\documentclass[%
 reprint,
 amsmath,amssymb,
 aps,
]{revtex4-2}

\usepackage{graphicx}% Include figure files
\usepackage{dcolumn}% Align table columns on decimal point
\usepackage{bm}% bold math
\usepackage{color}
\begin{document}

\preprint{APS/123-QED}

\title{
%Brain criticality revealed through nonadditive entropic analysis of electroencephalograms\\
%or\\
%Entropic analysis of electroencephalograms reveals brain criticality\\
%or\\
Brain criticality through nonadditive entropic analysis of electroencephalograms}% Force line breaks with \\

\author{ Henrique S.  Lima$^1$, Constantino Tsallis$^{1,2,3,4}$, and Dimitri M. Abramov$^5$}

\affiliation{ $^1$ Centro Brasileiro de Pesquisas Físicas, Rua Xavier Sigaud 150, Rio de Janeiro 22290-180, Brazil }
\affiliation{$^2$ Santa Fe Institute, 1399 Hyde Park Road, Santa Fe, NM 87501, USA}
\affiliation{$^3$ Complexity Science Hub Vienna, Metternichgasse 8, 1030 Vienna, Austria}
\affiliation{$^4$ Dipartimento di Fisica e Astronomia Ettore Majorana, University of Catania, Italy}
\affiliation{$^5$ Instituto Nacional da Saúde da Crianca, da Mulher e do Adolescente Fernandes Figueira. Fundação Oswaldo Cruz, Avenida Rui Barbosa 716, Flamengo, Rio de Janeiro 22250-020, Brazil.}

\date{\today}% It is always \today, today,

\begin{abstract}
On the grounds of nonadditive entropies -- appropriate for complex systems -- we investigate the 
%of segments 
electroencephalogram amplitudes of typical and ADHD children. 
The corresponding probability distributions
%of these segments 
%are well fitted 
%by a standard function of $q$-statistics: 
are $q$-Gaussians, i.e., $\rho(x) \propto 
%= \rho_0\, 
e_q^{-\beta x^2} \equiv [1+(q-1) \beta x^2]^{1/(1-q)}$, 
%where $\rho_0$ is the normalization constant,
where $(q,\beta)$ are, respectively, the entropic index characterizing complexity and the inverse width. 
We show that $q$ tends to monotonically vary with $\beta$ for both typical and ADHD subjects, thus revealing  critical behavior of the brain. Moreover, we verify that ADHD subjects have a higher complexity than the typical ones. Consistently, biomarkers for objective phychyatric diagnosis could emerge along this path.

\end{abstract}

\maketitle

%\section{Introduction}

%In recent years, advances in psychiatry and neuroscience have led to increasingly refined and segmented diagnostic criteria for neurodevelopmental and cognitive disorders. 

\textit{Introduction -} In recent years, numerous scientific advances have provided a growing understanding of the neurobiological correlates of mental disorders. However, there are still no specific markers that can support an objective diagnosis, which currently remains based on clinical and statistical descriptive criteria \cite{DSM2013}. Insufficient knowledge about the neurobiology of mental disorders still prevents the establishment of precision psychiatry, given the %criticality 
complexity of the brain \cite{Abramov2026}, fundamentally manifesting a state of criticality between great order and great disorder underlying self-organizing transitions to different collective functional states \cite{OByrneJerbi2022, Tian2022, Zimmern2020}.

%While this growing sophistication has improved clinical classification, it also raises a fundamental challenge: how to quantify and statistically characterize such disorders in a robust and objective manner. 

In particular, Attention Deficit Hyperactivity Disorder (ADHD) remains a condition whose % dynamical signatures are not yet fully understood from a quantitative perspective 
underlying mechanisms are not fully understood and, therefore, objective diagnosis remains ellusive \cite{Peterson2024, Koirala2024}.

The study of the complexity of the brain function and its criticality is an issue of clinical interest\cite{Abramov2026, Zimmern2020}, thus complexity measures appear as a possible path to solve the problem of accurately diagnosing mental disorders. Numerous evidence links pathological states and maladaptive processes (such as aging and mental disorders) to changes in the complexity of the dynamics of crucial systems such as the brain \cite{Abramov2026, Goldberger2002, Yang2013,  Chatterjee2017,Hadoush2019,Gu2022, Lau2022}. %AJEITAR AS REFERENCIAS.

A promising approach lies in the statistical analysis of neurophysiological signals such as electroencephalograms (EEGs), which encode rich information about brain activity across multiple temporal and spatial scales \cite{Niedermeyer2011}. Beyond traditional spectral and time-domain analyses, recent efforts have focused on identifying universal statistical features capable of distinguishing typical from atypical brain dynamics \cite{Lau2022}. 

Through statistical distributions of intervals between events in the EEG, which describe its temporal regularity, neural complexity has been studied in adults and children, both typical and with ADHD, in various functional brain states, using $q$-exponential functions emerging within non-extensive statistical mechanics \cite{Nosso2023, Nosso2024, Nosso2025}, which, grounded on nonadditive entropies, generalize the standard Boltzmann-Gibbs (BG) theory (recovered as the $q=1$ particular instance). In particular, the sensitivity of $q$-statistics in order to differentiate between typical children and those with ADHD has been shown \cite{Nosso2024}.

In this work, we investigate the statistical properties of EEG signal segments from both typical individuals and subjects diagnosed with ADHD. Our approach is based on the hypothesis that brain dynamics, as reflected in EEG signals, may exhibit non-Gaussian features that are not adequately captured by standard statistical frameworks.

To address this, we analyze the probability distributions of EEG segments within the context of nonextensive statistical mechanics. In particular, we consider the $q$-Gaussian distribution,
\begin{equation}
\rho(x) = \rho_0\, e_q^{-\beta x^2},
\end{equation}
where $\rho_0$ is a normalization constant, $q$ is the entropic index, and $\beta$ is a positive parameter inversely proportional to the width; 
%The $q$-exponential function is defined as 
$e_q^x\equiv[1+(1-q)x]_{+}^{1/(1-q)}$,  where  $[\dots]_{+}=[\dots]$ if $[\dots]$ is positive and zero otherwise ($e_1^x=e^x$)  \cite{Tsallis1988}. 

This framework provides a natural generalization of the Gaussian distribution and has been successfully applied to systems exhibiting long-range correlations, memory effects, and anomalous fluctuations \cite{TsallisBukman1996,LyraTsallis1998,LenziMalacarneMendes2003,TsallisGellMannSato2005, UmarovTsallisSteinberg2008}.

Within this setting, we aim to characterize the statistical behavior of EEG signals and to explore whether the parameters $(q,\beta)$ can serve as meaningful descriptors for distinguishing between typical and ADHD brain activities.
%\section{Model and methods}
\begin{figure*}[htb]
     \centering
     \includegraphics[width=0.45\linewidth]{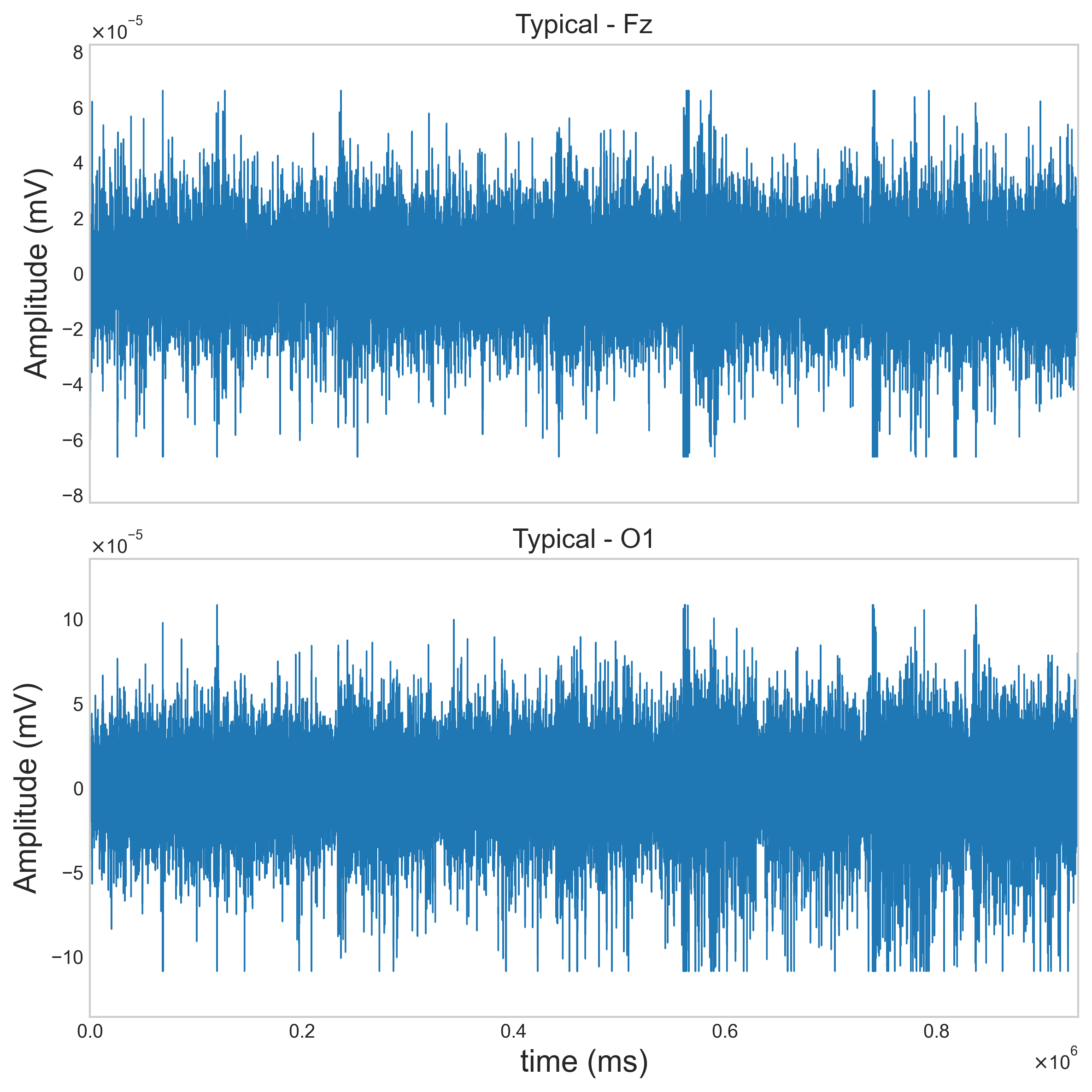}
     \includegraphics[width=0.45\linewidth]{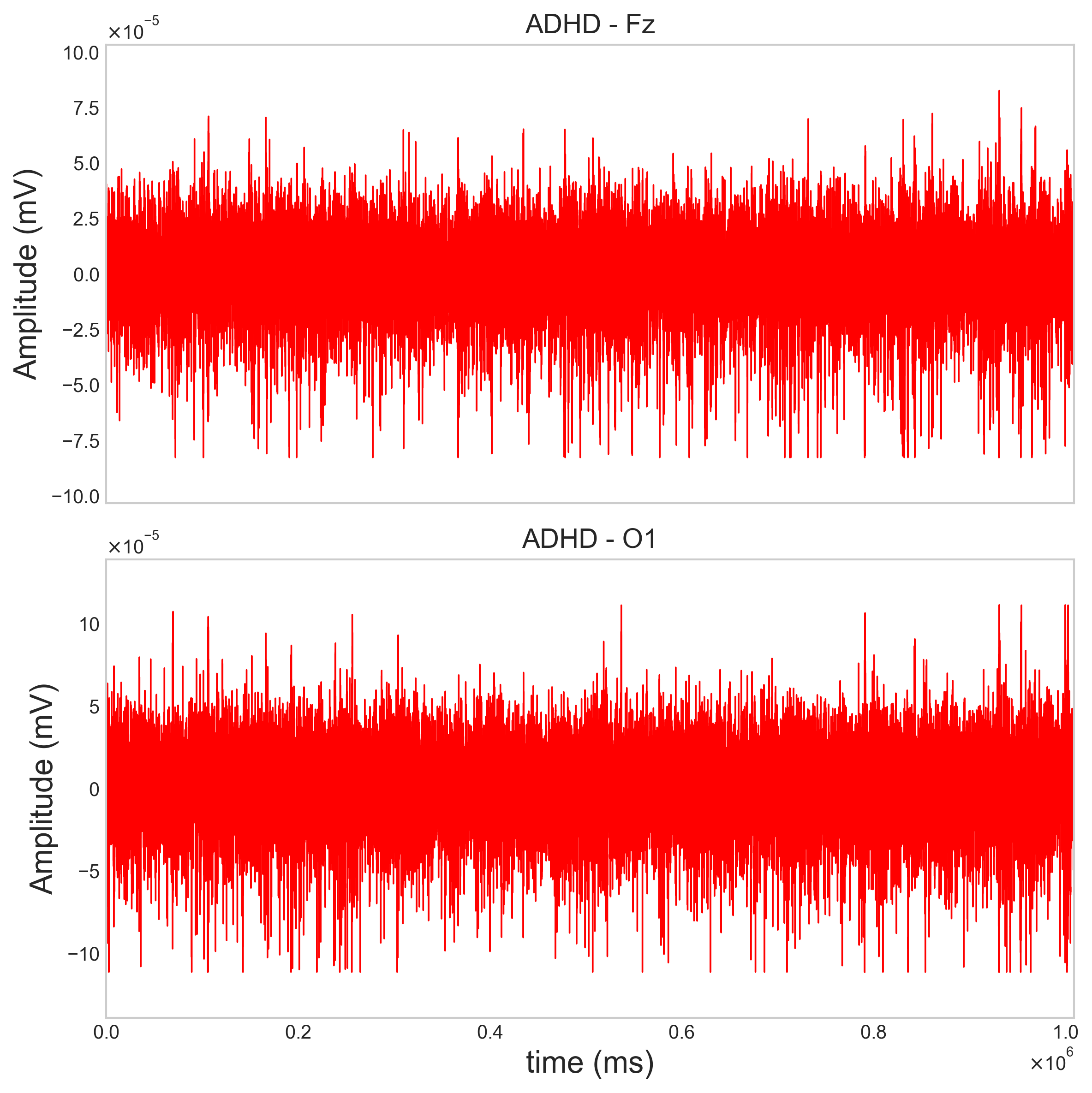}     
    \caption{Signal segments  for the Fz and O1 channels for four arbitrarily chosen patients.}
     \label{fig1}
 \end{figure*}
%\subsection{Data Preprocessing}

\textit{Model and methods} - We have analyzed, through a different property, the complexity of the same EEG signals recorded during Attention  Network Test (ANT) from 19 typical and 19 ADHD boys (11 to 13 years old) from our previous studies \cite{Nosso2024, Abramov2019}. The raw EEG signals were band-pass filtered between 0.5~Hz and 100.0~Hz and corrected for power-line interference using a notch filter at 60, 120, and 180~Hz. Each channel was zero-meaned along time, and extreme amplitude values were clipped using a \(z\)-score threshold of \(z = 5.0\). After this preprocessing, the signals were segmented into stationary time windows, and the amplitude fluctuations within each segment were normalized to define the variable \(x\) used in the subsequent statistical analysis. In Fig.\ref{fig1}   some arbitrarily chosen signal segments  are depicted. The amplitude fluctuations are very distinct from Gaussian noises showing large peaks around their mean values. These peaks are well known to be very common in systems characterized by strong nonlocal correlations, hence the theoretical approach through $q$-statistics is a priori plausible.
%\subsection{Statistical Modeling and Fitting}

For each preprocessed EEG segment, the empirical probability distribution of the fluctuations \(x\) was estimated via a normalized histogram, using a bin count adapted to the number of samples. The resulting probability density \(\rho(x)\) was then fitted to the \(q\)-Gaussian ansatz
\[
\rho(x) = \rho_0\, e_q^{-\beta x^2},
\]
by mapping the problem to a weighted linear regression. Specifically, we defined the transformed variable \(y = \ln_q\!\left(\rho / \rho_0\right)\) and performed a weighted least squares regression of \(y\) versus \(x^2\) through the origin, so that the fitted slope yields \(-\beta\). The corresponding entropic index \(q\) was determined by maximizing the coefficient of determination \(R^2\) of this linear relation over a wide interval \(q \in [q_\text{min}, q_\text{max}]\), using a combination of grid search and local refinement. The fitted parameters \((q,\beta)\) were recorded for each channel and subject, enabling subsequent statistical comparison between typical and ADHD groups.

%\section{Results}

%\begin{figure*}
%\centering
    %\includegraphics[width=0.8\linewidth]{f8t3t4.png}
    %\includegraphics[width=0.8\linewidth]{t5t6fz.png}
    %\includegraphics[width=0.56\linewidth]{czpz.png}
    %\caption{Plot of $q$ versus $1/\beta^{\gamma}$ for the channels F8, T3, T4, T5, T6, Fz, Cz, and Pz for all patients.}
    %\label{fig2b}
%\end{figure*}
\begin{figure*}
     \centering
     \includegraphics[width=0.95\linewidth]{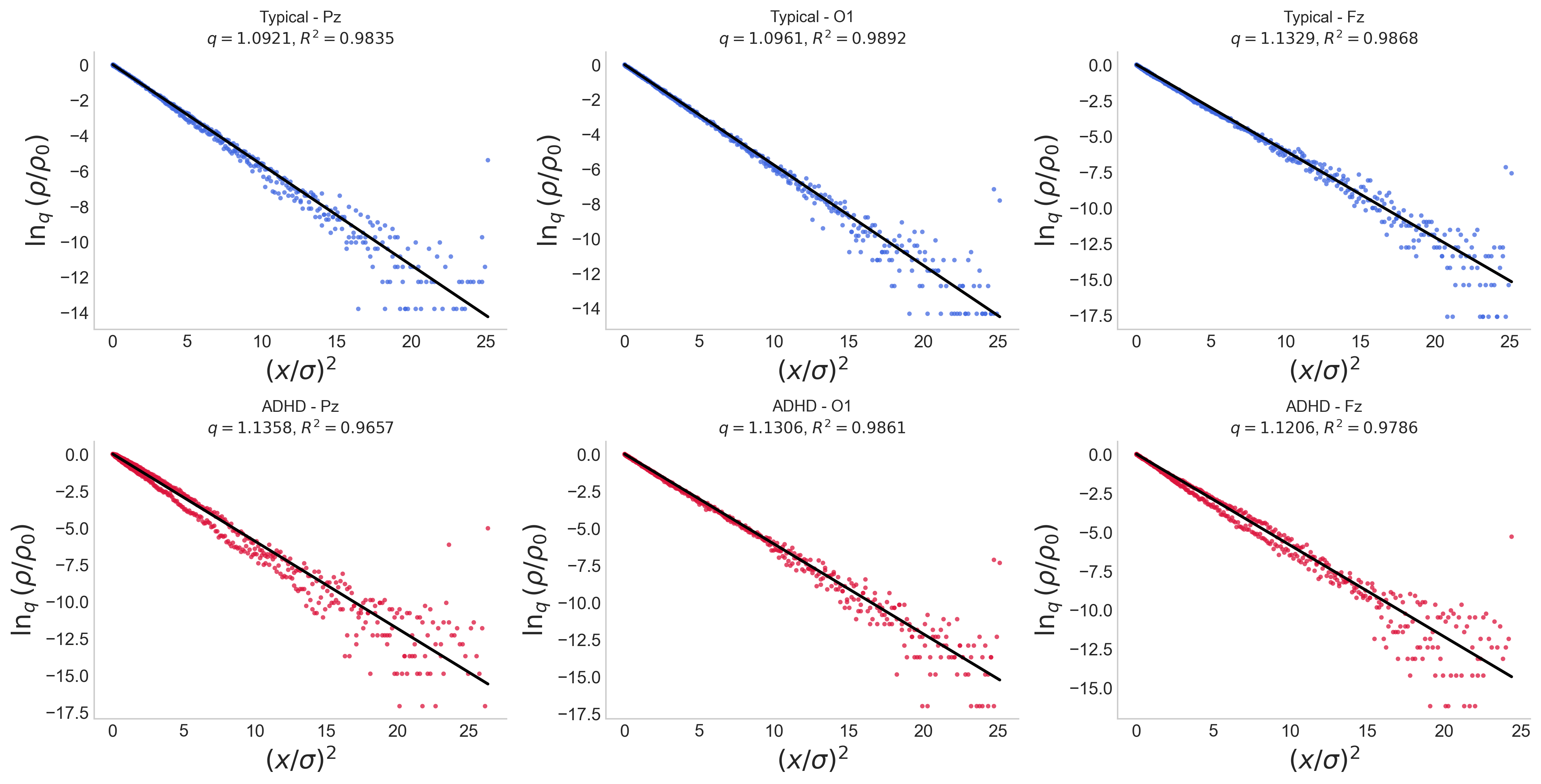}  
         \caption{Curves of $\ln_q(\rho/\rho_0)$ versus $(x/\sigma)^2$ for the Pz, O1, and Fz channels, where $\sigma$ is the standard deviation used to normalize the signal. The choices of the channel and the subject were arbitrary and made only for illustration. The dimensionless variable $x/\sigma$ was used only for better visualization. }
     \label{fig2}
 \end{figure*}

\begin{figure*}
    \centering
    \includegraphics[width=0.9\linewidth]{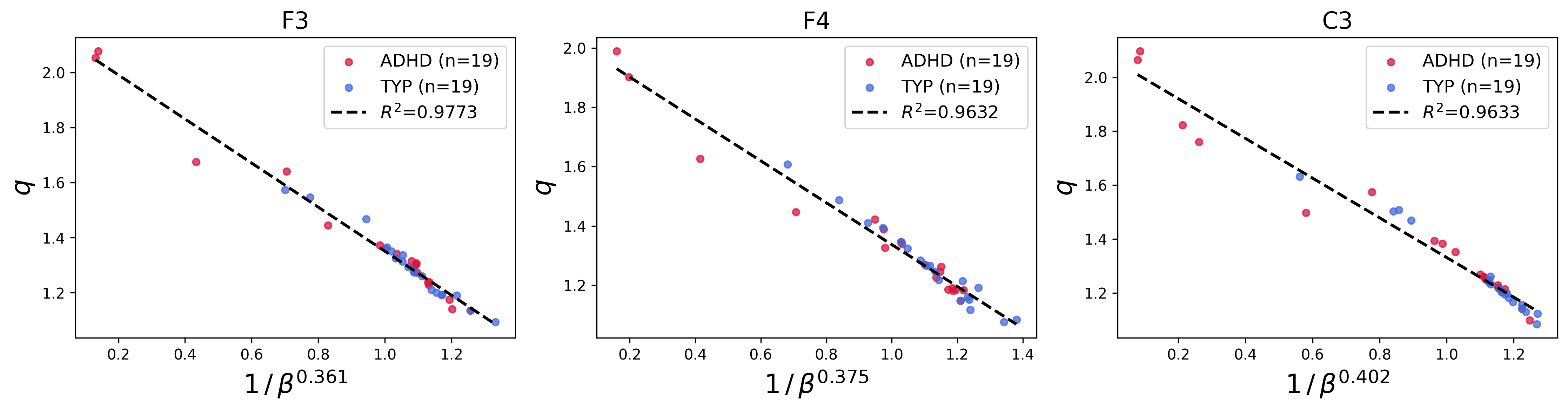}
    \includegraphics[width=0.9\linewidth]{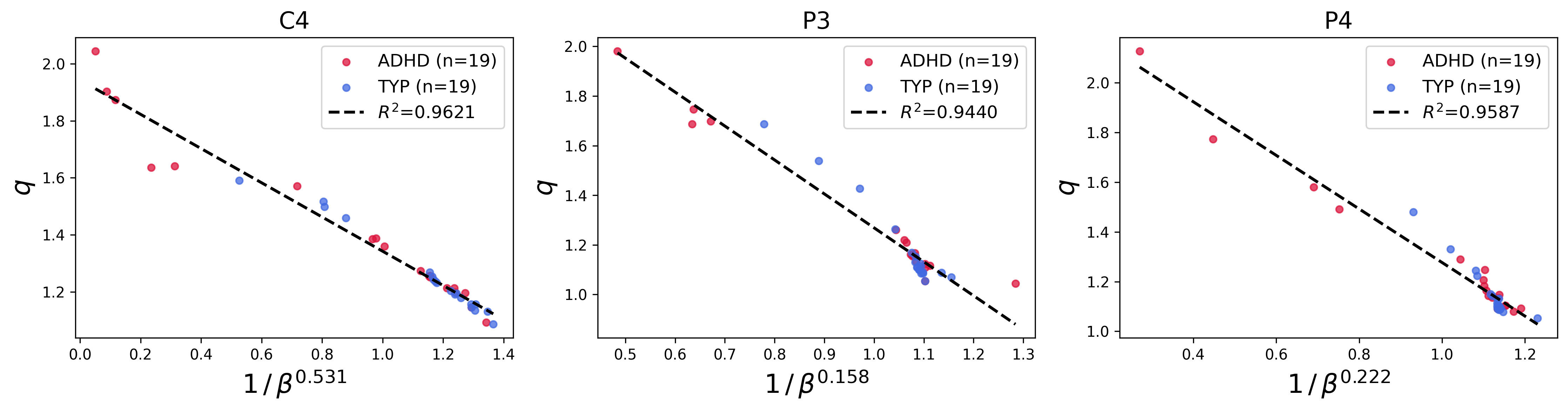}
    \includegraphics[width=0.9\linewidth]{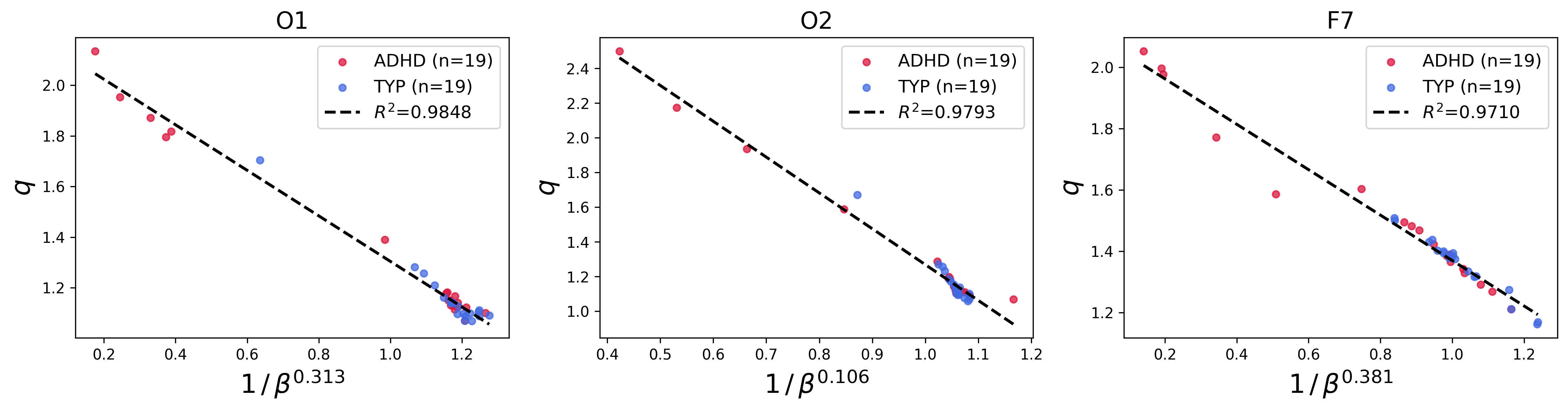}    
    \caption{Plot of $q$ versus $1/\beta^{\gamma}$ for the channels F3, F4, C3, C4, P3, P4, O1, O2, and F7 for all patients. Similar results have been obtained for the remaining channels.}
    \label{fig3}
\end{figure*}

\textit{Results} - In Fig.~\ref{fig2}, we arbitrarily illustrate some probability distributions in the scale $\ln_q(\rho/\rho_0)$ versus $(x/\sigma)^2$, which clearly exhibit linear behavior. This representation was intentionally chosen because $\ln_q(e_q^x)=x$; therefore, since $\ln_q(1)=0$, we obtain exactly $\ln_q(\rho/\rho_0)=\beta x^2$. 
%The variance $\sigma$ enters into the definition of the $\beta$ values. 

Table~\ref{typadhdmeanvalues} shows that the mean value of $q_{ADHD}$ for ADHD individuals is slightly larger than the mean value of $q_{typ}$ for typical ones. However, the standard deviation of $q_{ADHD}$ is also larger than that of $q_{typ}$, as are their relative deviations $\sigma_q/\langle q \rangle$. The same behavior occurs for $\beta^{-\gamma}$, where $\gamma$ is independently determined for each channel.

In Fig.~\ref{fig3}, we observe that $q=q(\beta)$ varies as an affine function of $1/\beta^\gamma$, namely $q_i(\beta)=c_i-1/\beta^{\gamma_i} \;\; (i=1,2, \dots, 17)$ , where the $(c_i,\gamma_i)$ values differ slightly for the various channels. It is known that the $q$-Gaussian distribution decays asymptotically as $\rho \sim A \, x^{-2/(q-1)}$, where $A$ is a constant. Consequently, the exponent $2/(q-1)$ becomes $2/(c_i-1+1/\beta^{\gamma_i})$. Notice that when $\beta \to \infty$, this yields $2/(c_i-1)$. From the results in Fig.~\ref{fig3}, we obtain that $q_i(\infty)=c_i$ lies between $1.9$ and $2.5$, i.e., $1.9<c_i<2.5$, which corresponds to an exponent approximately in the range $4/3 < 2/[q_i(\beta)-1] < 2.23$.

\begin{table*}
\centering
\includegraphics[width=0.7\linewidth]{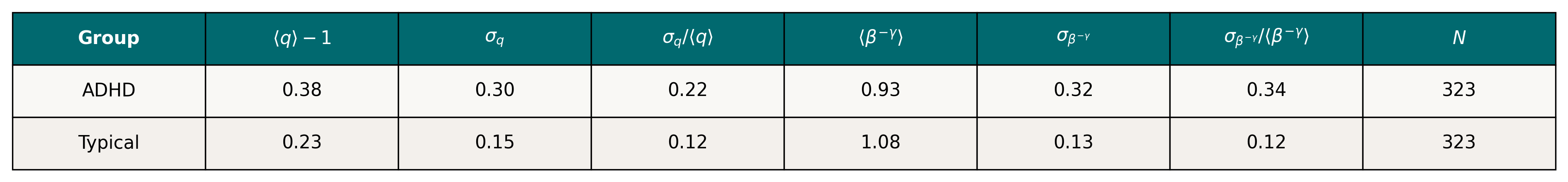}
\caption{Mean values of $q$ and $\beta^{-\gamma}$.}
\label{typadhdmeanvalues}
\end{table*}

\begin{figure*}
\centering
    \includegraphics[width=0.75\linewidth]{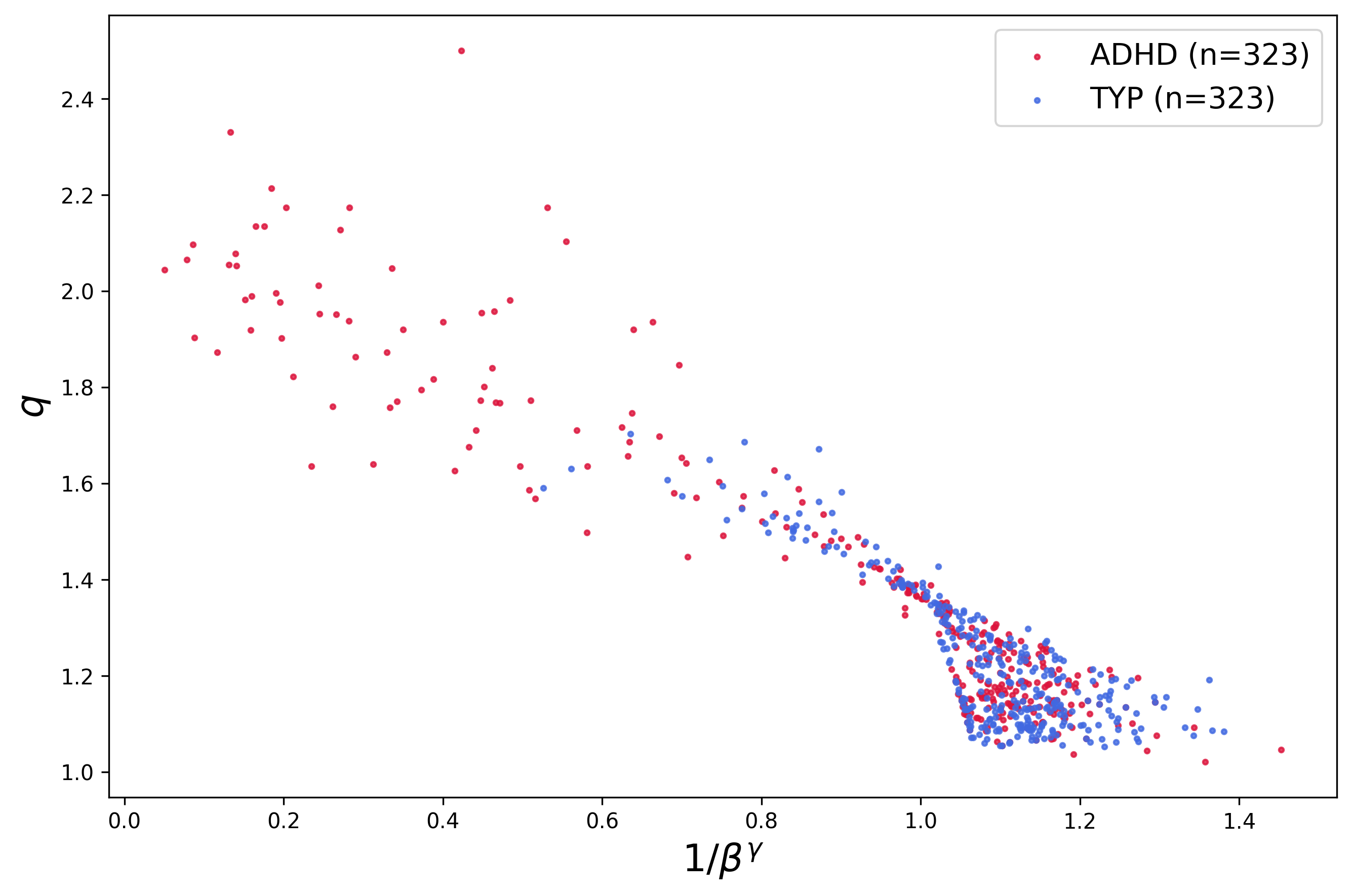}
    
    \caption{Plot of $q$ versus $1/\beta^\gamma$ for all 17 channels and all 38 patients. }
    \label{fig4}
    \end{figure*}

Despite the $\gamma$ values differing across channels for both typical and ADHD individuals, the graph of $q$ versus $\beta^{-\gamma}$ for all channels shows a clear collapse (see Fig. \ref{fig4}). This suggests two possible explanations: (i) ADHD and typical subjects belong to the same universality class, thus exhibiting no significant differentiation, or (ii) some of the 19 patients diagnosed with ADHD are in fact false positives. Explanation (i) is depicted in Figs.~\ref{fig3} and \ref{fig4}, where the criticality occurs along the same straight lines for both ADHD and typical results. This outcome suggests that they share the same exponents, and therefore belong to the same universality class.
 Explanation (ii) can be more clearly seen in Fig.~\ref{fig4}, where for all channels it is very difficult to differentiate the ADHD points from the typical ones. If explanation (i) were incorrect, we would expect to observe a cluster or at least some differentiation between patients, which does not occur in this case. Let us clarify that if (i) is correct, then (ii) might also hold true. However, if (ii) is correct, the lack of knowledge about the diagnosis does not allow any definitive conclusion regarding (i).

Table~\ref{typadhdchannels} presents the $\alpha$ values for each channel. Notice that the channels F3, O1, O2, and Fz have the highest $R^2$ values. Although F3, O1, and Fz share similar $\gamma$ values, the channel O2 exhibits an exponent $\gamma$ that is markedly different from those of the other channels, and it can be compared to the exponents of P3 and T6. The highest $\gamma$ value is associated with the C4 channel.

\begin{figure*}
\centering

\includegraphics[width=0.90\linewidth]{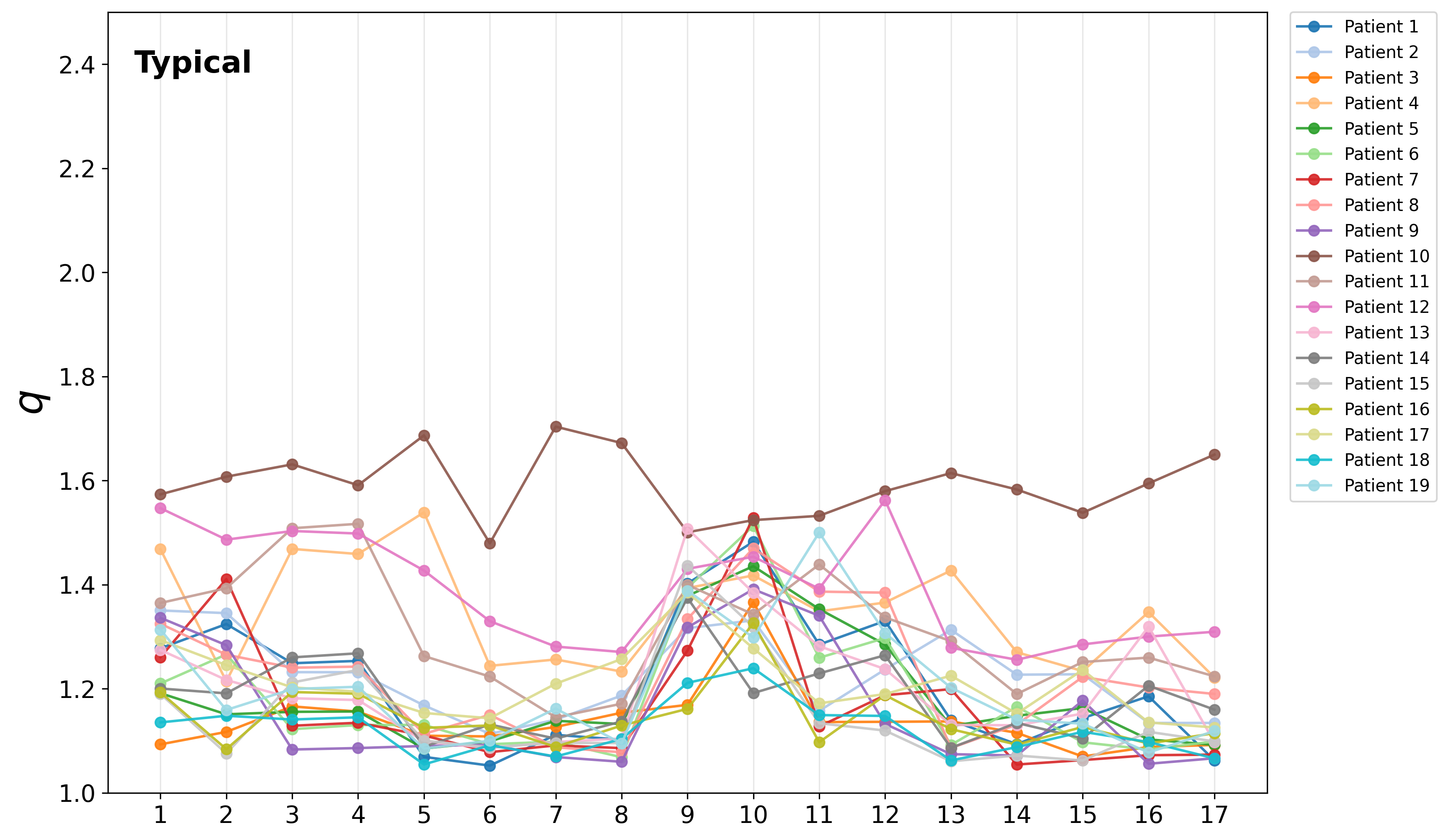}
    \includegraphics[width=0.90\linewidth]{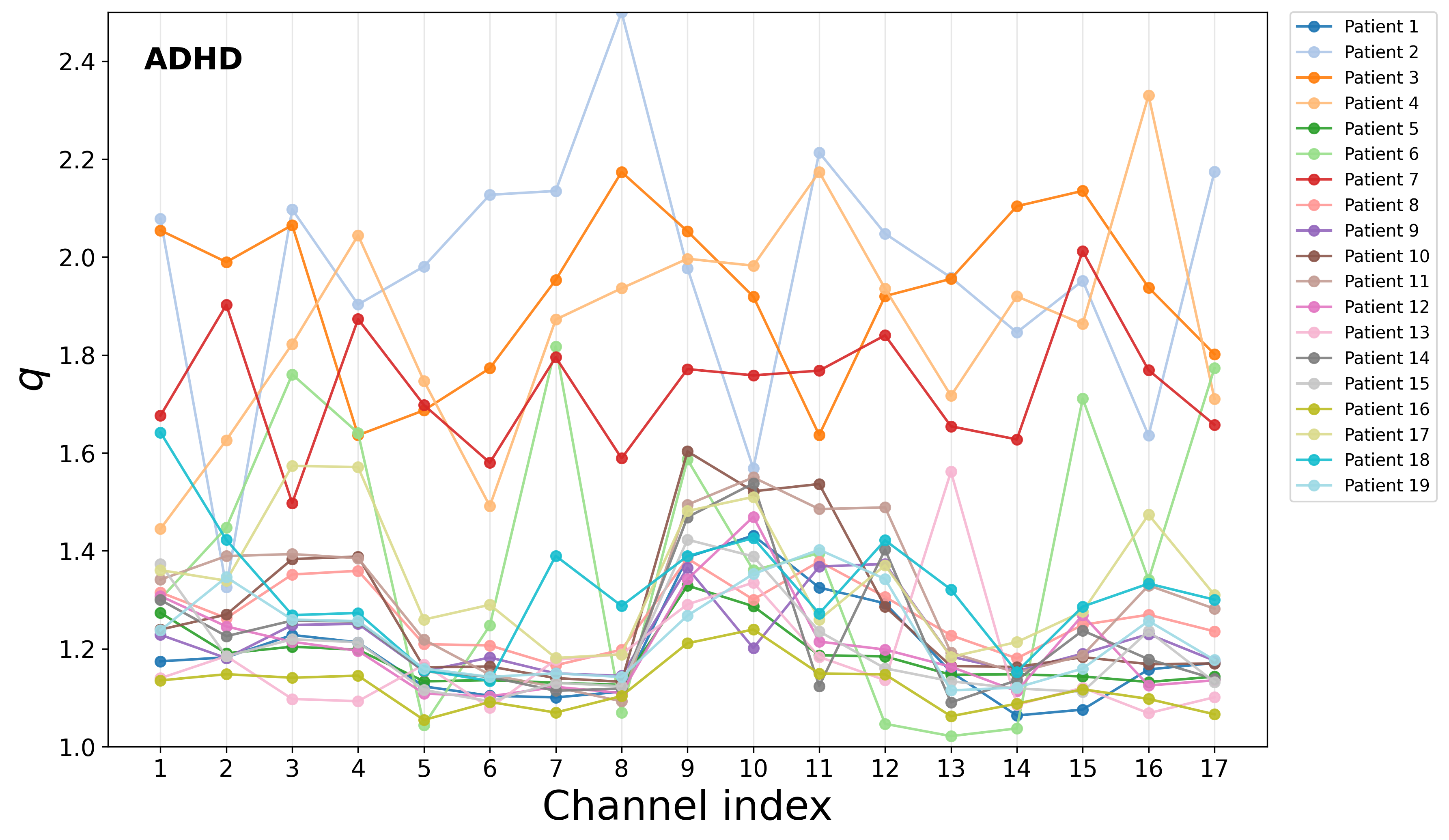}
        
    \caption{Plot of $q$ versus channel index for all 38 subjects. The channels are ordered as follows: F3, F4, C3, C4, P3, P4, O1, O2, F7, F8, T3, T4, T5, T6, Fz, Cz, Pz (17 channels in total). }
    \label{fig5}
    \end{figure*}

In Fig.~\ref{fig5}, we present the channel indices  $q$ for both typical and ADHD individuals.
For the typical group, no strong anomalies are observed. For the ADHD group, four patients exhibit channel indices whose $q$ values are sensibly larger than the values observed for typical patients. 
These results neatly exhibit that the ADHD behavior is associated with a complexity higher than the normal one.

\begin{table*}
\centering
\includegraphics[width=0.5\linewidth]{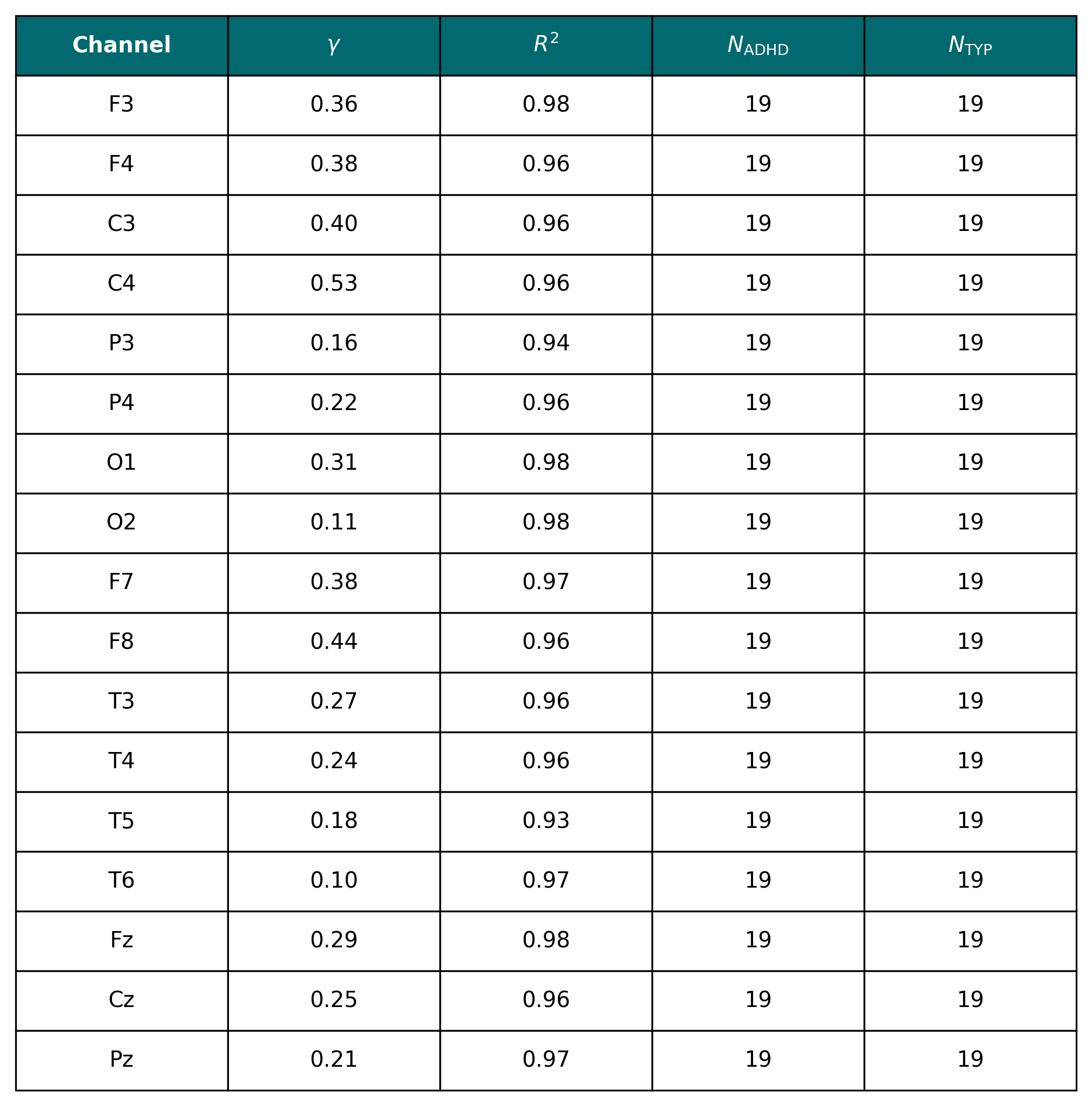}
\caption{Values of $\gamma$ corresponding to the channels  F (frontal), C (central), O (occipital), P (parietal), and T (temporal).}
\label{typadhdchannels}
\end{table*}
%\section{Discussion}

%\begin{figure*}
%\centering
%\includegraphics[width=0.6\linewidth]{meanqchversuspindextyp.png}  \includegraphics[width=0.6\linewidth]{meanqchversuspindexADHD.png}     
%\caption{Plot of  $\langle q\rangle_{ch}$ ($\langle\, .  \rangle_{ch}$ corresponds to the average over channels) versus patient index in ascending order versus channel index. }
%\label{fig3}
%\end{figure*}
    
\begin{figure*}
\centering

\includegraphics[width=0.7\linewidth]{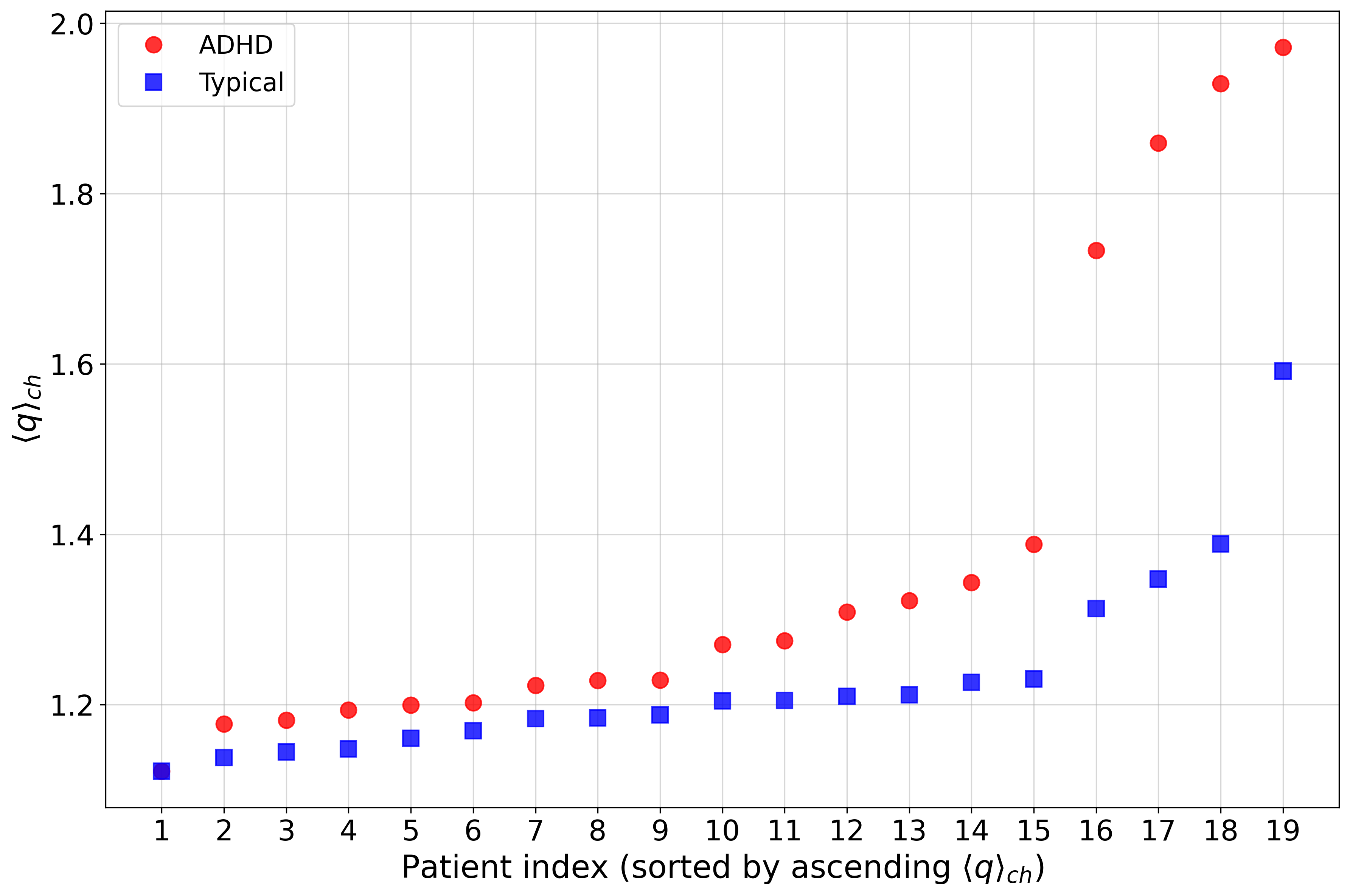}
        
    \caption{Plot of  $\langle q\rangle_{ch}$ (average over channels) versus patient index in ascending order. 
    The indices of the first ADHD and first typical patients coincide.}
    \label{fig6}
    \end{figure*}
    
%\section{Conclusions}

\textit{Final remarks} - In this work, we have analyzed the statistical properties of EEG signal segments from both typical individuals and subjects diagnosed with ADHD within the framework of nonextensive statistical mechanics. Our results show that the distributions of EEG fluctuations are well described by $q$-Gaussians, indicating the presence of nonlocal correlations which cause sensible deviations from standard Gaussian behavior.

A central finding is that the parameters $(q,\beta)$ are not independent, but instead exhibit a power-law-like  relationship for both groups. This scaling behavior suggests the existence of an underlying organizing principle, compatible with the emergence of critical-like dynamics in brain activity~\cite{OByrneJerbi2022, Tian2022}. Notably, although both typical and ADHD subjects follow similar functional trends, the corresponding parameter regimes differ, pointing to distinct dynamical signatures. A monotonic dependence of $q$ on $\beta$ has been encountered in various complex systems \cite{RuizMarcos, Grecoetal2020, Oliveiraetal2022}, and it is in variance with that corresponding to the distribution of delays in British railways (see Fig. 5 of \cite{BriggsBeck2007}).

These results support the idea that the $q$-generalized statistical approach provides a useful tool for capturing the complexity of neural signals \cite{Nosso2023, Nosso2024, Nosso2025, Abramov2026}. More broadly, our findings contribute to the ongoing effort to establish quantitative and physically grounded markers for mental disorders. In particular, the eight individuals whose values of $q$ distinctively differ from the others (see patients 16 to 19 in Fig. \ref{fig6}) may indicate limitations of current clinical diagnosis procedures.  
Future work may extend this approach by exploring larger datasets, different brain states or disorders, and potential connections with underlying neurophysiological mechanisms.

The present results in Fig. \ref{fig6} do not yet enable for individual diagnosis of ADHD excepting if $\langle q\rangle_{ch}$ is above say $1.6$. Moreover, the sudden increases of the indices in the last four patients of both ADHD and typical populations naturally suggest the need for further studies, preferentially in larger populations, in order to verify whether they indicate false positives or severe degrees of the ADHD behavior, or even the simultaneous incidence of other neurological anomalies.   

Last but not least,
it would undoubtedly be interesting to apply to girls the same approach that has been here applied to boys. Indeed, it is known that girls exhibit only mild evidence suggesting ADHD. There is, however, a longstanding controversy on whether the etiology is  behavioral ({\it masking}) or neurophysiological. The verification, or not, for girls of something similar to what is exhibited for boys in Fig. \ref{fig6}, could clarify this timely issue.

%\section*{Acknowledgments}
We acknowledge fruitful discussions with  L. B. C. de Fran\c{c}a, M. L. da Rocha, A. C. Tsallis and A. F. Tsallis. We also acknowledge partial financial support from CAPES, CNPq, FAPERJ (Brazilian agencies), {as well as FIOTEC/FIOCRUZ through the Research Incentive Program (IFF/FIOCRUZ).

\end{document}